\documentclass[11pt]{article}\textwidth 6.5in\textheight 51pc
\usepackage{amssymb}\usepackage[colorlinks]{hyperref}\usepackage{color}
\usepackage{mathrsfs}\usepackage[normalem]{ulem}\usepackage{amsthm}
\usepackage{amsmath}
\usepackage{amssymb}
\usepackage{verbatim}
\usepackage{setspace}
\usepackage{MnSymbol}
\usepackage{scrextend}
\usepackage{graphicx}
\usepackage{caption}

\newcommand\K{{\mathbf K}} \newcommand\I{{\mathbf I}}
  
\renewcommand\d{{\mathbf d}} 
\newcommand\Ks{\mathbf{Ks}} 
\newcommand\E{{\mathbf E}}

\newcommand\N{\mathbb{N}}\newcommand\BT{\Sigma}
\newcommand\FS{\BT^*}\newcommand\IS{\BT^\infty}
\newcommand\FIS{\BT^{*\infty}}

\newtheorem{thr}{Theorem} 
\newtheorem{lem}{Lemma}

\newtheorem{lmm}{Lemma}

\newtheorem{dff}{Definition}

\newcommand\floor[1]{{\lfloor#1\rfloor}}\newcommand\ceil[1]{{\lceil#1\rceil}}

\newcommand{\lea}{<^{+}}
\newcommand{\gea}{>^{+}}
\newcommand{\eqa}{=^{+}}

\newcommand{\lel}{<^{\log}}
\newcommand{\gel}{>^{\log}}
\newcommand{\eql}{=^{\log}}

\renewcommand\N{\mathbb{N}}
\renewcommand\BT{\Sigma}
\renewcommand\FS{\BT^*}
\renewcommand\IS{\BT^\infty}
\renewcommand\FIS{\BT^{*\infty}}

\renewcommand\i{\mathbf{i}}

\renewcommand\K{{\mathbf K}} 

\renewcommand\I{{\mathbf I}}

\renewcommand\d{{\mathbf d}}

\begin{document}

\author {Samuel Epstein\footnote{JP Theory Group. samepst@jptheorygroup.org}}
\title{\vspace*{-3pc} On the Kolmogorov Complexity of Binary Classifiers}
 \date{\today}\maketitle
\begin{abstract}
We provide tight upper and lower bounds on the expected  minimum Kolmogorov complexity of binary classifiers that are consistent with labeled samples. The expected size is not more than complexity of the target concept plus the conditional entropy of the labels given the sample.
 \end{abstract}
\section{Introduction}
This paper provides bounds on the Kolmogorov complexity of binary classifiers. In machine learning, classification is the task of learning a binary function $c$  from $\N$ to bits $\BT$. The learner is given a sample consisting of pairs $(x,b)$ for string $x$ and bit $b$ and outputs a binary classifier $h:\N\rightarrow\BT$ that should match $c$ as much as possible. Occam's razor says that "the simplest explanation is usually the best one." Simple hypothesis are resilient against overfitting to the sample data. In Section \ref{sec:rw}, we show some areas of machine learning which directly use the principle of Occam's razor. The question is, given a particular problem in machine learning, how simple can the hypotheses be?

In this paper, we provide tight upper and lower bounds of the length of the minimum description (a.k.a. Kolmogorov complexity, $\K$) of hypotheses that are consistent with the labelled sample data. 

We use a probabilistic model. The target concept is modeled by a random variable $\mathcal{X}$ with distribution $p$ over ordered lists of natural numbers. The random variable $\mathcal{Y}$ models the labels, and has a distribution over lists of bits, where the distribution of $\mathcal{X}\times\mathcal{Y}$ is $p(x,y)$ with conditional probability requirement $p(y|x)=\prod_{i=1..|x|}p(y_i|x_i)$. Each such $(x_i,y_i)$ is a labeled sample. A binary classifier $f$ is consistent with labelled samples $(x,y)$, if for all $i$, $f(x_i)=y_i$. Let $\Gamma(x,y)$ be the minimum Kolmogorov complexity of a classifier consistent with $(x,y)$. $\mathcal{H}(\mathcal{Y}|\mathcal{X})$ is the conditional entropy of $\mathcal{Y}$ given $\mathcal{X}$.\\

\noindent\textbf{Theorem.}
\begin{enumerate}
	\item $\mathcal{H}(\mathcal{Y}|\mathcal{X})\leq\E[\Gamma(\mathcal{X},\mathcal{Y})]\lel \mathcal{H}(\mathcal{Y}|\mathcal{X})+\K(p)$.
\item \textit{For each $c,b\in\N$, there exists random labeled samples $\mathcal{X}\times \mathcal{Y}$ with distribution $p$, such that, up to precision $O(\log cb)$, $\E[\Gamma(\mathcal{X},\mathcal{Y})]=b+c$, $\mathcal{H}(\mathcal{Y}|\mathcal{X})=b$, and $\K(p)= c$.	 }
 \end{enumerate}\newpage

\section{Related Work}
\label{sec:rw}
There are many places the machine learning literature where Occam's razor is used. The goal of the extremely successful Minimum Description Length principle (see \cite{GrunwaldMyPi05}) is to find the most succinct hypothesis to model the data. This involves choosing a hypothesis $H$ from a set of candidates that minimizes the length of the ``code'' for $H$, $L(H)$, plus the length of the code describing the data $D$ given $H$, denoted $L(D|H)$. One difference between this paper and MDL is that MDL uses computable codes $L$, where we use the non-computable minimum program size.

The Occam Learning Algorithm \cite{BlumerEhHaWa87} shows that a concept class is learnable if one can succinctly describe the training data. A concept $c$ is a binary function over a finite set of strings. A concept class $C$ is a set of such functions.
The  Occam Learning Algorithm takes in $m$ labelled samples $s$ of a concept $c$ (a.k.a binary function) with encoding length $\mathrm{Size}(c)$ and returns a hypothesis $h$ consistent with $c$ on $s$ with $\mathrm{Size}(h) \leq (n\times\mathrm{Size}(c))^\alpha m^\beta$, for some $\alpha\geq 0$ and $0\leq \beta<1$, where $n$ is max length among the samples $s$. If a concept class has an Occam Algorithm, then it is PAC-learnable \cite{Valiant84} and has a finite VC dimension \cite{Vapnik98}, which means that the concept class can be efficiently learned. Thus succinct representation of the hypothesis is connected to learnability of a target concept. 

The structural risk minimization principle in statistical learning theory \cite{Vapnik98}  looks for the optimal relationship between the quality of the approximation of the target concept by a hypothesis and the complexity class which the hypothesis is in (a.k.a VC dimension). Thus simply described hypothesis will be more valued than complicated ones.
\subsection*{Algorithmic Information Theory}
In \cite{Epstein21}, it was shown that the minimum description length of a binary classifier consistent with $n$ samples is less than $n$ plus the amount of information that the samples have with the halting sequence. Theorem \ref{thr:clopen} generalizes this result to clopen sets and computable measures.

The study of Kolmogorov complexity originated from the work of~\cite{Kolmogorov65}. The canonical self-delimiting form of Kolmogorov complexity was introduced in~\cite{ZvonkinLe70} and treated later in~\cite{Chaitin75}. More information about the history of the concepts used in this paper can be found the textbook~\cite{LiVi08}. 

Theorem \ref{thr:clopen} in this paper is an inequality including the mutual information of the encoding of a finite set with the halting sequence. A history of the origin of the mutual information of a string with the halting sequence can be found in~\cite{VereshchaginVi04v2}.

A string is stochastic if it is typical of a simple elementary probability distribution. A string is typical of a probability measure if it has a low deficiency of randomness. The notion of the deficiency of randomness with respect to a measure follows from the work of~\cite{Shen83}, and also studied in~\cite{KolmogorovUs87,Vyugin87,Shen99}. Aspects involving stochastic objects were studied in~\cite{Shen83,Shen99,Vyugin87,Vyugin99}. \newpage

\section{Conventions}
\label{sec:conv}
Let  $\BT$, $\FS$, and $\IS$ be the sets of bits, finite strings, and infinite strings. We use $\langle x\rangle$ to represent a self delimiting code for $x\in\FS$, with $\langle x\rangle=1^{\|x\|}0x$.  For sets $C\subseteq\IS$ and $D\subseteq\FS$, $C\unlhd D =\{x:x\in D, \Gamma_x\subseteq C\}$. A clopen set in the Cantor space is a finite union og intervals. For clopen set $C$, $\langle C\rangle=\langle \{x : \Gamma_x\textrm{ is maximal in } S\}\rangle$. The indicator function of a mathematical statement $A$ is denoted by $[A]$, where if $A$ is true then $[A]=1$, otherwise $[A]=0$.

For positive real functions $f$ the terms $\lea f$, $\gea f$, and $\eqa f$ represent $<f+O(1)$, $>f-O(1)$, and $=f\pm O(1)$, respectively. For nonnegative real function $f$ the terms $\lel f$, $\gel f$ and $\eql f$ represent $< f+O(\log(f+1))$, $>f-O(\log(f+1))$, and $=\pm O(\log(f+1))$, respectively.

A probability measure $Q$ over $\N$  is elementary if $|\{a:Q(a)>0\}|<\infty$ and $\mathrm{Range}(Q)$ consists of all rationals. Elementary measures $Q$ can be encoded into finite strings $\langle Q\rangle$.

We use a universal prefix free algorithm $U$, where we say $U_\alpha(x)=y$ if $U$, on main input $x$ and auxiliary input $\alpha$, outputs $y$. We define Kolmogorov complexity with respect to $U$, with for $x\in\FS$, $y\in\FIS$, $\K(x/y)=\min\{\|p\|:U_y(p)=x\}$. By the chain rule, $\K(x,y)\eqa\K(x)+\K(y/x,\K(x))$. The halting sequence $\mathcal{H}\in\IS$ is the unique infinite sequence where $\mathcal{H}[i]=[U(i)\textrm{ halts}]$. The information that $x\in\FS$ has about $\mathcal{H}$, conditional to $y\in\FS\cup\IS$, is $\I(x;\mathcal{H}/y)=\K(x|y)-\K(x/\langle y,\mathcal{H}\rangle)$. The Kolmogorov complexity of an infinite sequence $\alpha\in\IS$ is the size of the smallest input to $U$ which will output, without halting, $\alpha$ on the output tape.

This paper uses notions of stochasticity in the field of algorithmic statistics \cite{VereshchaginSh17}. A string $x$ is stochastic, i.e. has a low $\Ks(x)$ score if it is typical of a simple probability distribution. The deficiency of randomness function of a string $x$ with respect to an elementary probability measure $P$ is $\d(x|P)=\floor{-\log P(x)}-\K(x|\langle P\rangle)$. 
\begin{dff}[Stochasticity]$ $
	For $x,y\in\FS$,\\  $\Ks(x)=\min \{\K(P)+3\log \max\{\d(x|P),1\}: P\textrm { is an elementary probability measure}\}$.
\end{dff}

\section{Results}

Each binary classifier can be represented as an infinite sequence in the natural way. The following theorem is a statement about measures and clopen sets. It may be of independent interest, as it generalizes Theorem 6 from \cite{Epstein21}.
\begin{thr}$ $\\
	\label{thr:clopen}
	For clopen set $C\subseteq\IS$, computable measure $S$,
	 $\min_{\alpha\in C}\K(\alpha)\lel  -\log S(C)+\I(C;\mathcal{H})+O(\K(S))$.
\end{thr}
\begin{prf}
Let $s=\ceil{-\log S(C)}$.
We remove consideration of the complexity terms of $S$ and $s$ in the proof because of the size of the error terms of the theorem. Let $P$ be an elementary probability measure that realizes $\Ks(\langle C\rangle)$.  Let $n$ be the maximum length of members of  finite sets encoded in the support of $P$. More formally,  $n=\max \{\|x\|:x \in W\subset\FS, \langle W\rangle\in\mathrm{Supp}(P)\}$. The max term can be used because $P$ is elementary, and thus has a finite support. 
	
The randomness deficiency of $S$ with respect to $P$ is  $d=\max\{\d( C|P),1\}$. Let $c\in\N$ be a constant solely dependent on the universal Turing machine $U$ to be determined later. Let $\kappa$ be a probability measure over lists $L$ of $cd2^s$ strings of length $n$, where $\kappa(L)=\prod_{i=1}^{cd2^{s}}S(L_i)$. Let $\i(W,L)$ be an indicator function over sets $W\subset\FS$ and lists $L\subseteq\BT^n$,  with $\i(W,L)=[S(W)\geq 2^{-s},W\unlhd L=\emptyset]$.
	\begin{align*}
	\E_{L\sim\kappa}\E_{\langle W\rangle\sim P}[\i( W,L)] &\leq \sum_{ \textrm{clopen }W\subseteq\IS}P(\langle W\rangle)\left(1-2^{-s}\right)^{cd2^s}\leq e^{-2^{-s} cd2^{s}}=e^{-cd}.
	\end{align*}
	Thus there exists a list $L$ of $cd2^s$ strings such that $\E_{\langle W\rangle\sim P}[\i(W,L)]<e^{-cd}$. This $L$ can be found with brute force search,  with $\K(L|c,d,P)=O(1)$.  Using $L$, we can define the following $P$-test, $t(W)=e^{cd}\i(W,L)$, with $\sum_WP(W)t(W)\leq 1$.
	 It must be that $C\unlhd L\neq\emptyset$, otherwise $t$ will give $C$ a high score, with  
	 $t(C)=e^{cd}$. This causes the following contradiction for large enough $c$ solely dependent on the universal Turing machine $U$, with
	\begin{align*}
	\K(C|c,d,\langle P\rangle) &< -\log t_L(C)P(\langle C\rangle)+O(1)\\
	\K(C|c,d,\langle P\rangle) &< -\log P(\langle C\rangle)-(\lg e)cd+O(1)\\
	(\lg e)cd &< -\log P(\langle C\rangle)-\K(C|\langle P\rangle)+\K(d,c)+O(1)\\
	(\lg e)cd &< d + \K(d,c)+O(1).
	\end{align*}
	We roll $c$ into the additive constants of the rest of the proof. So there exists $x \in C\unlhd L$, with
	\begin{align*}
	\K(x)&\lea \log \big|L\big|+\K(L)\\
	&\lea \log \big|L\big|+\K(d,P)\\
	&\lea \log d+s + \K(d)+\K(P)\\
	&\lea s+\Ks(\langle C\rangle).
	\end{align*}
	Since $x\in C\unlhd L$, $\Gamma_x\subseteq C$. Thus there is a program $g$ that outputs $x$ and then an infinite sequence of 0's. Since $x0^\infty\in C$ and $\|g\|\lea \K(x)$,
	\begin{align*}
	\min_{\alpha\in C}\K(\alpha)\leq\|g\|\lea \K(x)\lea s+\Ks(\langle C\rangle).
	\end{align*}
Using Lemma 10 in \cite{Epstein21}, which states $\Ks(x)\lel\I(x;\mathcal{H})$, we get the final form of the proof, 
\begin{align}
	\min_{\alpha\in C}\K(\alpha)\lel s+\I(C;\mathcal{H})+O(\K(S)).
\end{align}
\end{prf}
The following lemma is perhaps surprising because it shows that the $\I(\cdot;\mathcal{H})$ terms in inequalities can be removed by averaging over a computable probability.
\begin{lmm}
	\label{lmm:prodh}
	For computable probability $p$, $\sum_xp(x)\I(x;\mathcal{H})\lea \K(p)$.
\end{lmm}
\begin{prf}
	This follows from Theorem 3.1.3 in \cite{Gacs21}, and we will reproduce its arguments. Since $\K(x/\mathcal{H})$ is the length of a self delimiting code,
	 $$\sum_xp(x)\K(x/\mathcal{H})\geq \mathcal{H}(p),$$
	 where $\mathcal{H}(p)$ is the entropy of $p$. Furthermore, for all $x\in\FS$, $\K(x)\lea-\log p(x)+\K(p)$. Therefore $$\sum_xp(x)\K(x)\lea \sum_xp(x)(-\log p(x))+\K(p)\lea \mathcal{H}(p)+\K(p).$$
	 So
	 \begin{align*}
	 \sum_xp(x)\I(x;\mathcal{H})
	 &= \sum_xp(x)\left(\K(x)-\K(x/\mathcal{H})\right)
	 \lea \mathcal{H}(p)+\K(p)-\sum_xp(x)\K(x/\mathcal{H})
	\lea \K(p).
	 \end{align*}
	 
\end{prf}
The following theorem is the main result of the theorem. It essentially involves averaging the inequality of Theorem \ref{thr:clopen} over the target probability. The $\I(\cdot;\mathcal{H})$ term vanishes due to Lemma \ref{lmm:prodh}. The theorem is slightly better than the simpler statement in the introduction. The terms in the theorem are defined in the introduction.

\begin{thr}
	\label{thr:main}
	$\E[\Gamma(\mathcal{X},\mathcal{Y})]< \mathcal{H}(\mathcal{Y}|\mathcal{X})+\K(p) +O(\log \mathcal{H}(\mathcal{Y}|\mathcal{X}))$.
\end{thr}	
\begin{prf}
	Binary classifiers are identified by infinite sequences $\alpha\in\IS$. We define the computable measure $S$, where $S(x)=\prod_{n=1..|x|}p(x_n|n)$, where $\K(S|p)=O(1)$. Let $\{(x_i,y_i)\}$ be a set of labelled samples and we define clopen set $C_{x,y}=\{\alpha:\alpha\in\IS,\alpha[x_i]=y_i\}$. Then $S(C_{x,y})=p(y|x)$. By Theorem \ref{thr:clopen}, relativized to $p$,
	\begin{align}
	\nonumber
	\min_{\alpha\in C_{x,y}}\K(\alpha|p)&\lel -\log S(C_{x,y})+\I(C_{x,y};\mathcal{H}|p)+O(\K(S|p))\\
		&\lel -\log S(C_{x,y})+\I(C_{x,y};\mathcal{H}|p)\\
		\nonumber
	&\lel -\log p(y|x)+\I(C_{x,y};\mathcal{H}|p)\\
		\label{eq:1}
	\sum_{x,y}p(x,y) \min_{\alpha\in C_{x,y}}\K(\alpha|p)&\lel \sum_{x,y}p(x,y)(-\log p(y|x))+\sum_{x,y}p(x,y)\I(C_{x,y};\mathcal{H}|p).
	\end{align}
	Applying Lemma \ref{lmm:prodh} relative to $p$, we get
		\begin{align}
		\label{eq:2}
		\sum_{x,y}p(x,y)\I(C_{x,y};\mathcal{H}|p)&\lea \K(p|p)=O(1).
	\end{align}
	Combining equations \ref{eq:1} and \ref{eq:2},
	\begin{align*}
	\sum_{x,y}p(x,y) \min_{\alpha\in C_{x,y}}\K(\alpha|p)&< \sum_{x,y}p(x,y)(-\log p(y|x))+ O\left( \sum_{x,y}p(x,y)\log(-\log p(y|x) )\right)\\
	&< \sum_{x,y}p(x,y)(-\log p(y|x))+ O\left(\log \sum_{x,y}p(x,y)(-\log p(y|x) )\right)\\
	\E[\Gamma(\mathcal{X},\mathcal{Y})]-\K(p)&< \mathcal{H}(\mathcal{Y}|\mathcal{X})+O(\log \mathcal{H}(\mathcal{Y}|\mathcal{X}))\\
	\E[\Gamma(\mathcal{X},\mathcal{Y})] &< \mathcal{H}(\mathcal{Y}|\mathcal{X})+\K(p)+O(\log \mathcal{H}(\mathcal{Y}|\mathcal{X})).
	\end{align*}
\end{prf}

The following theorems provides lower bounds to the Kolmogorov complexity of classifiers. 

\begin{thr}
	$\mathcal{H}(\mathcal{Y}|\mathcal{X})\leq \E[\Gamma(\mathcal{X},\mathcal{Y})]$
\end{thr}
\begin{prf}
	$\E[\Gamma(\mathcal{X},\mathcal{Y})]=\sum_xp(x)\sum_yp(y|x)\Gamma(x,y)$. For a fixed $x$, ranged over $y$, $\Gamma(x,y)$ represents the length of a self-delimiting code. Due to properties of conditional entropy, 	$\sum_xp(x)\sum_yp(y|x)\Gamma(x,y)\geq \sum_xp(x)\sum_yp(y|x)(-\log p(y|x))=\mathcal{H}(\mathcal{Y}|\mathcal{X})$.
\end{prf}\\

\begin{thr}
	For each $c,b\in\N$, there exists random labeled samples $\mathcal{X}\times \mathcal{Y}$ with distribution $p$, such that, up to precision $O(\log cb)$,
	\begin{enumerate}
		\item $\mathcal{H}(\mathcal{Y}|\mathcal{X})=b$,
		\item $\K(p)= c$, 
		\item $\E[\Gamma(\mathcal{X},\mathcal{Y})]=b+c$.	 
	\end{enumerate}
\end{thr}
\begin{prf}
	We ignore all $O(\log cd)$ terms. So equality $=$ is equivalent to $=\pm O(\log cd)$. We define a probability $p(x,y)$ over the first $n=2c+2b+2$ numbers and corresponding bits. Thus we can describe $p$ as a probability measure over strings of size $n$, making sure to maintain $p$'s conditional probability restriction described in the introduction.
	
	Let $z\in\BT^c$ be a random string of size $c$, with $c\lea \K(z)$. For all strings $w\in\BT^b$ of size $b$, $p(\langle z\rangle\langle w\rangle)=2^{-b}$, with $\|\langle z\rangle\langle w\rangle\|=n$. $\mathcal{H}(\mathcal{Y}|\mathcal{X})=-\sum_{w\in\BT^b}2^{-b}(\log p(\langle w\rangle\langle z\rangle))=-\sum_{w\in\BT^b}2^{-b}(\log 2^{-b})=b$. Furthermore $\K(p)=c$.
	
	The infinite sequence  $\alpha=\langle w\rangle\langle z\rangle 0^\infty$ realizes $\Gamma(\langle w\rangle \langle z\rangle )$ up to an additive constant for each $w\in\BT^b$. Thus $\K(\alpha)= \K(z,w)$.  
 $\E[\Gamma(\mathcal{X},\mathcal{Y})]=2^{-b}\sum_{w\in\BT^b}\K(\langle z\rangle\langle w\rangle)=\K(z)+2^{-b}\sum_{w\in\BT^b}\K(w/z,\K(z))$. Using  Theorem 3.1.3 in \cite{Gacs21} conditioned on $\langle z,\K(z)\rangle$, we get that $\sum_{w\in\BT^b}2^{-b}\K(w/z,\K(z))=\mathcal{H}(\mathcal{U}_b)\pm \K(b/z,\K(z))=b$, where $\mathcal{U}_b$ is the uniform measure over strings of size $b$. So $\E[\Gamma(\mathcal{X},\mathcal{Y})]=\K(z)+b=b+c$.
\end{prf}
\section{Discussion}
The results of this paper are for binary classifiers that are completely consistent with the sample data. One area of research is looking into the description length of classifiers that have a small classification error. 
In general case, the bounds of Theorem \ref{thr:main} cannot be improved upon. However, we hypothesize that there exist interesting sets of models $p$ whose expected classifier description length is much smaller than $\K(p)$. Another open research area is the intersection of algorithmic information theory with other areas of statistical learning theory, including density estimation and regression. 
%\bibliographystyle{alpha} 
%\bibliography{refnotes}

\begin{thebibliography}{BEHW87}
	
	\bibitem[BEHW87]{BlumerEhHaWa87}
	A.~Blumer, A.~Ehrenfeucht, D.~Haussler, and Warmuth.
	\newblock Occam's razor.
	\newblock {\em Information processing letters}, 24(6):377--380, 1987.
	
	\bibitem[Cha75]{Chaitin75}
	G.~J. Chaitin.
	\newblock {A Theory of Program Size Formally Identical to Information Theory}.
	\newblock {\em Journal of the ACM}, 22(3):329--340, 1975.
	
	\bibitem[Eps21]{Epstein21}
	Samuel Epstein.
	\newblock All sampling methods produce outliers.
	\newblock {\em IEEE Transactions on Information Theory}, 67(11):7568--7578,
	2021.
	
	\bibitem[G\'21]{Gacs21}
	Peter G\'{a}cs.
	\newblock Lecture notes on descriptional complexity and randomness.
	\newblock {\em CoRR}, abs/2105.04704, 2021.
	
	\bibitem[GMP05]{GrunwaldMyPi05}
	P.~Grunwald, I.~Myung, and M.~Pitt.
	\newblock {\em {Advances in minimum description length theory and
			applications}}.
	\newblock MIT Press, Cambridge, MA, USA, 2005.
	
	\bibitem[Kol65]{Kolmogorov65}
	A.~N. Kolmogorov.
	\newblock {Three approaches to the quantitative definition of information.}
	\newblock {\em {Problems in Information Transmission}}, 1:1--7, 1965.
	
	\bibitem[KU87]{KolmogorovUs87}
	A.~N. Kolmogorov and V.~A. Uspensky.
	\newblock {Algorithms and Randomness}.
	\newblock {\em {SIAM Theory of Probability and Its Applications}},
	32(3):389--412, 1987.
	
	\bibitem[LV08]{LiVi08}
	M.~Li and P.~Vit{\'a}nyi.
	\newblock {\em An {I}ntroduction to {K}olmogorov {C}omplexity and {I}ts
		{A}pplications}.
	\newblock Springer Publishing Company, Incorporated, 3 edition, 2008.
	
	\bibitem[She83]{Shen83}
	A.~Shen.
	\newblock {The concept of (alpha,beta)-stochasticity in the Kolmogorov sense,
		and its properties.}
	\newblock {\em {Soviet Mathematics Doklady}}, 28(1):295--299, 1983.
	
	\bibitem[She99]{Shen99}
	A.~Shen.
	\newblock {Discussion on Kolmogorov Complexity and Statistical Analysis}.
	\newblock {\em The Computer Journal}, 42(4):340--342, 1999.
	
	\bibitem[Val84]{Valiant84}
	L.~Valiant.
	\newblock A theory of the learnable.
	\newblock {\em Commun. ACM}, 27(11):1134–1142, 1984.
	
	\bibitem[Vap98]{Vapnik98}
	V.~Vapnik.
	\newblock {\em Statistical Learning Theory}.
	\newblock Wiley-Interscience, Hoboken, NJ, 1998.
	
	\bibitem[VS17]{VereshchaginSh17}
	Nikolay~K. Vereshchagin and Alexander Shen.
	\newblock Algorithmic statistics: Forty years later.
	\newblock In {\em Computability and Complexity}, pages 669--737, 2017.
	
	\bibitem[VV04]{VereshchaginVi04v2}
	N.~Vereshchagin and P.~Vit{\'a}nyi.
	\newblock {Kolmogorov's Structure Functions and Model Selection}.
	\newblock {\em IEEE Transactions on Information Theory}, 50(12):3265 -- 3290,
	2004.
	
	\bibitem[V'Y87]{Vyugin87}
	V.V. V'Yugin.
	\newblock {On Randomness Defect of a Finite Object Relative to Measures with
		Given Complexity Bounds}.
	\newblock {\em {SIAM Theory of Probability and Its Applications}}, 32:558--563,
	1987.
	
	\bibitem[V'Y99]{Vyugin99}
	V.V. V'Yugin.
	\newblock Algorithmic complexity and stochastic properties of finite binary
	sequences.
	\newblock {\em {The Computer Journal}}, 42:294--317, 1999.
	
	\bibitem[ZL70]{ZvonkinLe70}
	A.~K. Zvonkin and L.~A. Levin.
	\newblock The complexity of finite objects and the development of the concepts
	of information and randomness by means of the theory of algorithms.
	\newblock {\em Russian Math. Surveys}, page~11, 1970.
	
\end{thebibliography}

\end{document}